\documentclass[aps,amsmath,amssymb,showpacs,twocolumn,prl]{revtex4-1}
\usepackage{hyperref}
\usepackage{epsfig}
\usepackage{subfigure}
\usepackage{braket}
\usepackage{color}
\usepackage{verbatim}

\newcommand{\pardiff}[2]{\frac{\partial{#1}}{\partial{#2}}}
\newcommand{\ad}{a^{\dagger}}
\newcommand{\bd}{b^{\dagger}}
\newcommand{\Gu}{\Gamma_{\uparrow}}
\newcommand{\Gd}{\Gamma_{\downarrow}}
\newcommand{\tGu}{\tilde{\Gamma}_{\uparrow}}
\newcommand{\tGd}{\tilde{\Gamma}_{\downarrow}}

\newcommand{\modelabel}{m}
\newcommand{\nl}{n_\modelabel}

\begin{document}

\title{Nonequilibrium Model of Photon Condensation}

\author{Peter Kirton}
\affiliation{SUPA, School of Physics and Astronomy, University of St Andrews, St Andrews, KY16 9SS, United Kingdom}
\author{Jonathan Keeling}
\affiliation{SUPA, School of Physics and Astronomy, University of St Andrews, St Andrews, KY16 9SS, United Kingdom}
\date{\today}

\begin{abstract}
  We develop a nonequilibrium model of
  condensation and lasing of photons in a dye filled microcavity. We
  examine in detail the nature of the thermalization process induced
  by absorption and emission of photons by the dye molecules, and
  investigate when the photons are able to reach a thermal equilibrium
  Bose-Einstein distribution.  At low temperatures, or large cavity
  losses, the absorption and emission rates are too small to allow the
  photons to reach thermal equilibrium and the behavior becomes more
  like that of a conventional laser.
\end{abstract}

\maketitle

Bose-Einstein condensation (BEC) has been observed in a wide variety
of systems, from ultracold atomic gases~\cite{Anderson1995,
  Davis1995} to quasiparticles in solid state systems such as
polaritons~\cite{Kasprzak2006,Balili2007, Deng2010,Carusotto2012},
excitons~\cite{High2012}, and magnons~\cite{Demokritov2006}.  Recently
experiments have shown convincing evidence of a Bose-Einstein
distribution, and macroscopic occupation of the lowest mode for
a gas of photons confined in a dye-filled optical
microcavity~\cite{Klaers2010b, Klaers2010c, Klaers2011a,
  Schmitt2012a}.  In these experiments, the thermal equilibrium
distribution of photons arises because of phonon dressing of the
absorption and emission by the dye molecules, and the rapid
thermalization of rovibrational modes of the dye molecules by their
collisions with the solvent.  This leads to the accumulation of low-energy photons, closely following a Bose-Einstein distribution, as is
clearly seen experimentally~\cite{Klaers2010c}.

Such a system is very closely related to a dye
laser~\cite{Schafer1990}, but differs in the near-thermal emission
spectrum that is observed below and near the threshold density and in the fact
that the macroscopic population occurs at the minimum energy mode of the cavity and is not
related to the gain maximum of the dye~\cite{Klaers2010c}. There are
also close connections to microlasers~\cite{Walther2006}. However
microlasers, having strong coupling between the gain medium and cavity, display thresholdless lasing~\cite{Rice1994}.  In contrast, the
observed behavior in the photon condensate~\cite{Klaers2010c} is that
there is a sharp threshold which occurs far below inversion.

In the context of polariton condensation~\cite{Kasprzak2006,
  Balili2007, Deng2010,Carusotto2012} there has been much
debate~\cite{Butov2007,Snoke2008} about the extent to which the lack
of true thermal equilibrium in experiments means the system
should be called a condensate or a laser.  However, various calculations for
polaritons, from quantum
kinetics~\cite{doan05:prb, *Doan2006, *Doan2008,Kasprzak2008a} to
Schwinger-Keldysh path integrals~\cite{Szymanska2006,*Keeling2013},
have found a relatively smooth crossover between behavior typical of a
laser, and that typical of an equilibrium condensate.  Both lasers and
condensates involve a spontaneous phase-symmetry breaking, and a
transition to a macroscopically occupied mode, and so their connection
has long been recognized~\cite{haken75}.  The photon condensate system
provides a further example of a system in which the distinction
between Bose condensation and lasing must be carefully examined.

The nature of the thermalization process in the photon condensate
differs significantly from that found in other systems which exhibit
BEC. There are no direct photon--photon interactions in the cavity and
the thermal Bose-Einstein distribution seen in this system can be
understood as arising from the combination of asymmetry between
absorption and emission (the Kennard-Stepanov relation~\cite{Kennard1918,
  *Kennard1926, *Stepanov1957}) and the retrapping of fluorescence.
This mechanism and the presence of dissipation (loss) raises similar
questions to those raised for polaritons: Can the observed behavior
be understood as an exotic form of lasing? What features distinguish a
Bose-Einstein condensate from an exotic laser?  To address these
questions, we show that, starting from a model of \emph{stimulated
  emission}, i.e.\ that of a modified laser, we can describe the observed
\emph{Bose-Einstein distribution} of light.  For experimental
parameters (low losses), we find that the above threshold state is
practically indistinguishable from the
``textbook''~\cite{pitaevskii03} condensate of a noninteracting Bose
gas.  Significant deviations from the thermal behavior occur if the cavity
losses increase, and a crossover toward more standard lasing is
observed.  Our results therefore show that, even in an open system,
stimulated emission can produce a momentum distribution
indistinguishable from that arising in thermodynamic equilibrium.

Previous theoretical work has attempted to produce models of this
system from the point of view of \textit{equilibrium} statistical
mechanics~\cite{Klaers2012a, Sob'yanin2012}, while other work has examined
the emergence of phase coherence in a BEC where particles interact
through an intermediate medium~\cite{Snoke2012}.  We aim instead to
provide a general \emph{nonequilibrium} framework for understanding
the steady state properties of the photons, taking into account the
pump and decay processes.  This allows one to understand how these
compete with the thermalization process, and control when the system
behaves like a laser or like a condensate.

\begin{figure}
  \centering
  \includegraphics[width=2.9 in]{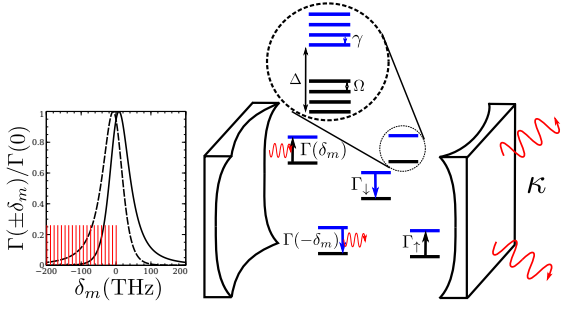}
  \caption{(Color online) Cartoon of the system showing the decay
    processes included in Eq.\ \eqref{eqn:ME}. The zoomed in view shows
    the energy level structure of the dye molecules. The graph shows the
    characteristic behavior of the emission rate, $\Gamma(-\delta_\modelabel)$
    (dashed line) and the absorption rate, $\Gamma(\delta_\modelabel)$ (solid
    line) described in the main text. The vertical (red) lines show the
    typical spacing of the photon modes confined in the
    cavity.}\label{fig1}
\end{figure}

A schematic diagram of our model is shown in Fig.~\ref{fig1}.
This consists of a photon modes, labeled $\modelabel$,  with creation operators
$a^\dagger_\modelabel$, confined in the optical cavity coupled to a
single electronic transition of the dye molecules. Each dye molecule, indexed by the label $i$,
is represented as a two-level system, with the Pauli matrix
$\sigma_i$ and splitting $\Delta$ between ground and excited levels.
These levels are dressed by ladders of rovibrational states, which can
be thought of as an on-site phonon~\cite{Schafer1990}, described by
operators $b_i, b^\dagger_i$. The level scheme is shown in
Fig.~\ref{fig1}.  The Hamiltonian is thus
\begin{multline}
  \label{eqn:H}
  H=\sum_{\modelabel}\omega_\modelabel\ad_\modelabel a_\modelabel
  +\sum_i\frac{\Delta}{2}\sigma_i^z
  +\Omega\left(\bd_ib_i+\sqrt{S}\sigma_i^z(b_i+\bd_i)\right)
  \\
  +
  g\sum_{\modelabel,i}\left(a_\modelabel\sigma_i^++\ad_\modelabel\sigma_i^-\right),
\end{multline}
using units such that $\hbar=k_B=1$. As in the experiment~\cite{Klaers2010c}, we consider photon modes in a
two-dimensional harmonic potential (arising from the curvature of the
mirrors).  We therefore take regularly spaced oscillator levels
$\omega_{\modelabel} = \omega_0 + \modelabel \epsilon$, having a
degeneracy $g_\modelabel$ given by $g_\modelabel=\modelabel+1$.  The
lowest frequency $\omega_0$ is the ``cavity cutoff.'' If in
equilibrium, condensation would lead to a macroscopic occupation
of this mode. In the following
we quote frequencies relative to the molecular splitting $\Delta$ and
thus introduce $\delta_\modelabel=\omega_\modelabel-\Delta$. 
Since the light-matter coupling is small compared to optical frequencies, we
assume a Jaynes-Cummings coupling, with frequency independent coupling
strength $g$.  The vibrational mode spacing is $\Omega$, and the
interaction between electronic and vibrational states is given by the
Huang-Rhys parameter $S$, which characterizes the
difference in phonon displacement between the ground and excited
states.  The parameter values we use, corresponding to the
experiment~\cite{Klaers2010c}, are given in the figure captions.

To model the open system, we must add additional loss processes and
external pumping.  We include the loss of cavity photons with rate
$\kappa$, assumed independent of the photon frequency, and a rate $\Gd$
describing fluorescence of the dye molecules into noncavity modes.
To balance these losses we include pumping with rate $\Gu$.
These processes may all be described by standard Markovian Lindblad
terms, as there is no significant thermal occupation of relevant
photon modes outside the cavity.  The localized vibrational modes also
undergo incoherent relaxation, due to scattering off of solvent
molecules.  This is modeled as a relaxation rate $\gamma$ toward a
thermal equilibrium state at temperature $T$.  These processes cannot
be described by Markovian loss rates, as this cannot describe
thermalization of the radiation~\cite{Ford1996,Lax2000}. Below we
describe an alternate approach to include these processes.

If the coupling to phonons $S$ is reasonably strong, then multiphonon
effects will be important in describing the thermalization processes.
These can be captured by making a polaron transformation $H\to
U^\dagger H U$, where $U=\exp[\sum_i\sqrt{S}\sigma_i^z(b_i-\bd_i)]$.
Since the coupling of molecules to the optical modes is weak, we then
treat the dynamics perturbatively in $g$ while keeping all orders of
$S$.  Working in the interaction picture, and expanding the Liouville
equation to second order in $g$, one may then trace out the degrees of
freedom associated with the vibrational mode and its damping.  The
resulting equation of motion then contains Lindblad terms which cause
simultaneous transitions in both the photon field and the dressed
molecule~\cite{Marthaler2011}.  These processes then describe the emission
(absorption) of photons into (from) the cavity, as shown schematically
in Fig.\ \ref{fig1}. Including all processes, the resulting master
equation for the photon-molecule system is
\begin{multline}
  \label{eqn:ME}
  \dot{\rho} =-i[H_0,\rho] -\sum_{i,\modelabel}\left\{
    \frac{\kappa}{2}\mathcal{L}[a_\modelabel]+\frac{\Gu}{2}\mathcal{L}[\sigma_i^+]+\frac{\Gd}{2}\mathcal{L}[\sigma_i^-]\right.\\\left.+\frac{\Gamma(-\delta_\modelabel)}{2}\mathcal{L}[\ad_\modelabel\sigma^-_i]+\frac{\Gamma(\delta_\modelabel)}{2}\mathcal{L}[a_\modelabel\sigma^+_i]\right\}\rho.
\end{multline}
Here, the system Hamiltonian is $H_0 = \sum_{\modelabel,i} \tilde\delta_\modelabel
\ad_\modelabel a_\modelabel+ \eta_\modelabel\ad_\modelabel a_\modelabel\sigma_i^+\sigma_i^-$ and $\mathcal{L}[X]\rho=\{X^\dagger X, \rho\} - 2 X
\rho X^\dagger$ is the usual Lindblad term.  The phonon assisted
emission and absorption rates, along with the Lamb shifts, are related to the function~\cite{Marthaler2011},
\begin{equation} \label{eqn:Gdelta}
	K(\omega)=g^2\int_{0}^\infty dt f(t){\rm e}^{-(\Gu+\Gd)|t|/2}{\rm e}^{-i\omega t},
\end{equation}
where $f(t)$ is
a correlation function of polaron operators~\cite{Wilson-Rae2002b,
  McCutcheon2011b, Marthaler2011} given by,
\begin{equation}
	f(t)=\exp\left[-\frac{2S\gamma}{\pi}\int_{-\infty}^\infty d\nu\frac{2\sin^2\frac{\nu t}{2}\coth\frac{\beta\nu}{2}+i\sin\nu t}{(\Omega-\nu)^2+\frac{\gamma^2}{4}}\right].
\end{equation}
The absorption and emission rates in the master equation are simply given by $\Gamma(\omega)=2{\rm Re}[K(\omega)]$ while the energy shifts in the Hamiltonian are $\eta_\modelabel={\rm Im}[K(-\delta_\modelabel)-K(\delta_\modelabel)]$ and $\tilde \delta_\modelabel =\delta_\modelabel+{\rm Im}[K(\delta_\modelabel)]$. These Lamb shifts do not affect the populations at order $g^2$; truncating at this order is valid below threshold in weak coupling. In the following we focus on the behavior below and at threshold and thus set $\eta_\modelabel=0$ and $\tilde \delta_\modelabel=\delta_\modelabel$~\cite{Marthaler2011}.
An illustration of the decay rates, as a function of detuning, is shown in
Fig.\ \ref{fig1}.  We note that for frequencies where the rates $\Gu,
\Gd$ can be ignored in Eq.\ \eqref{eqn:Gdelta}, the vibration induced
emission and absorption rates are related by a Boltzmann factor
$\Gamma(\delta)={\rm e}^{\beta\delta}\Gamma(-\delta)$\footnote{This
  expression arises due to a Kubo-Martin-Schwinger
  relation~\cite{Kubo1957, *Martin1959} between the time domain rates
  $f(t)=f(-t-i\beta)$.} with $\beta$ corresponding to the phonon
(solvent) temperature, thus satisfying the Kennard-Stepanov relation
between absorption and emission~\cite{Kennard1918, *Kennard1926,
  *Stepanov1957}.  At large frequencies $\Gamma(\omega)$ ceases to
obey this relation because the incoherent pumping process corresponds
to coupling to a white noise (i.e., infinite temperature) 
bath~\cite{Lax2000}.

~\citet{Marthaler2011} considered this kind of master equation as a
route to lasing without inversion in circuit QED.  The same mechanism
they proposed also applies for the photon condensate, allowing
coherent emission far below inversion: If
$\Gamma(-\delta_\modelabel)>\Gamma(\delta_\modelabel)$ then the asymmetry in
emission and absorption induced by thermalization with the phonons
allows net gain without inversion. For lasing to occur significantly
below the inversion point we require $\delta_0 \ll - T$ so that
the asymmetry of the absorption and emission rates is sufficiently
large.  As we will discuss below, the same conditions lead to
thermalized lasing, as long as the relevant emission rates
$\Gamma(-\delta_\modelabel)$ are large enough to overcome the losses from the cavity.

We can use the master equation, Eq.\ \eqref{eqn:ME}, to derive a
semiclassical rate equation for the population of each photon mode
which, after adiabatically eliminating the $N$ molecular degrees of
freedom, is
 \begin{equation} \label{eqn:dndt}
	\pardiff{\nl}{t}=-\kappa\nl+N\frac{\Gamma(-\delta_\modelabel)(\nl+1)\tGu-\Gamma(\delta_\modelabel)\nl\tGd}{\tilde \Gu+\tGd},
\end{equation}
where we define
$\tGu=\Gu+\sum_{\modelabel}g_{\modelabel}\Gamma(\delta_{\modelabel})\nl$
and
$\tGd=\Gd+\sum_{\modelabel}g_{\modelabel}\Gamma(-\delta_{\modelabel})(\nl+1)$.
We can then use the steady state of this expression in combination
with the rates from Eq.\ \eqref{eqn:Gdelta} to calculate the photon
population in each mode, $g_\modelabel n_\modelabel$.  Note that the
$n_\modelabel+1$ term in the emission process corresponds to the
trapping of spontaneous fluorescence from the dye.  When combined
with the Kennard-Stepanov relation discussed above, this means that
in the equilibrium limit, $\kappa,\Gu,\Gd\to0$, the stationary solution
to this equation is an equilibrium Bose-Einstein distribution.  i.e.\
$(n_m+1)/n_m = e^{\beta \delta_m} \tGd/ \tGu$.  Far below
threshold, where $n_m \ll 1$, this corresponds to an effective
chemical potential $\mu_{\text{eff}} = T \ln(\Gu/\Gd)$.  When
occupation is not negligible, the dependence of $\tGd$ and $\tGu$ on $n_m$ implies a self-consistency condition on $\mu$, so that
$\mu \to \delta_0$ as pumping increases.  This equilibrium limit is
discussed in Ref.~\cite{Klaers2012a}. Applying our rate equations above threshold the model continues to predict noninteracting BEC behavior; i.e., photons accumulate in the lowest energy mode and the thermal tail saturates. This is in contrast to the experiment~\cite{Klaers2010c} where the population in the tail continues to grow as expected for a trapped, interacting BEC~\cite{Tammuz2011}. However, as noted above, beyond threshold terms of order $g^4$ and above must be retained, describing thermalization of a BEC in which the interactions are
mediated by scattering~\cite{Tammuz2011, Wouters2012}.

 \begin{figure}
 \centering
\includegraphics[width=1.6in]{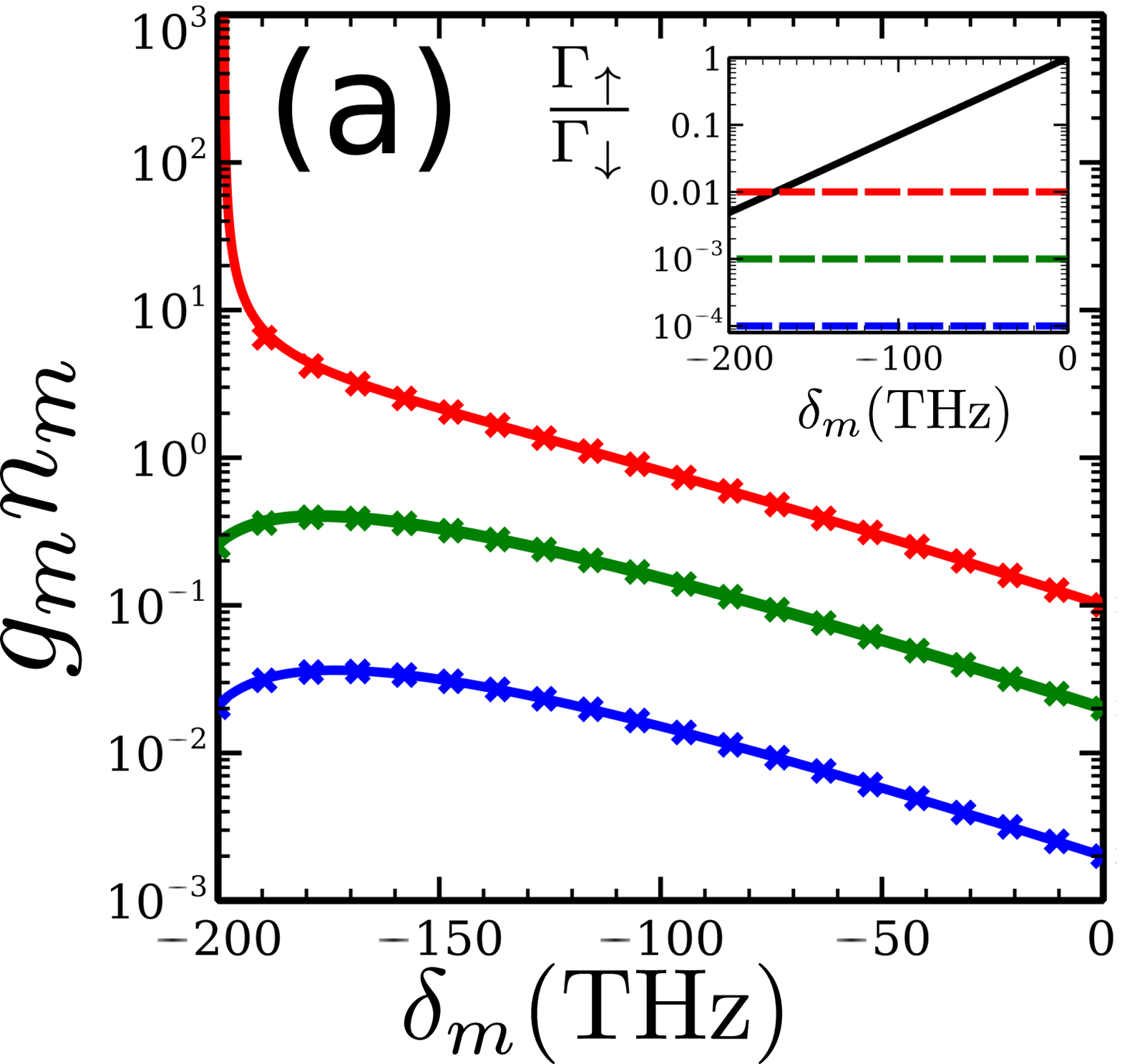}
\includegraphics[width=1.6in]{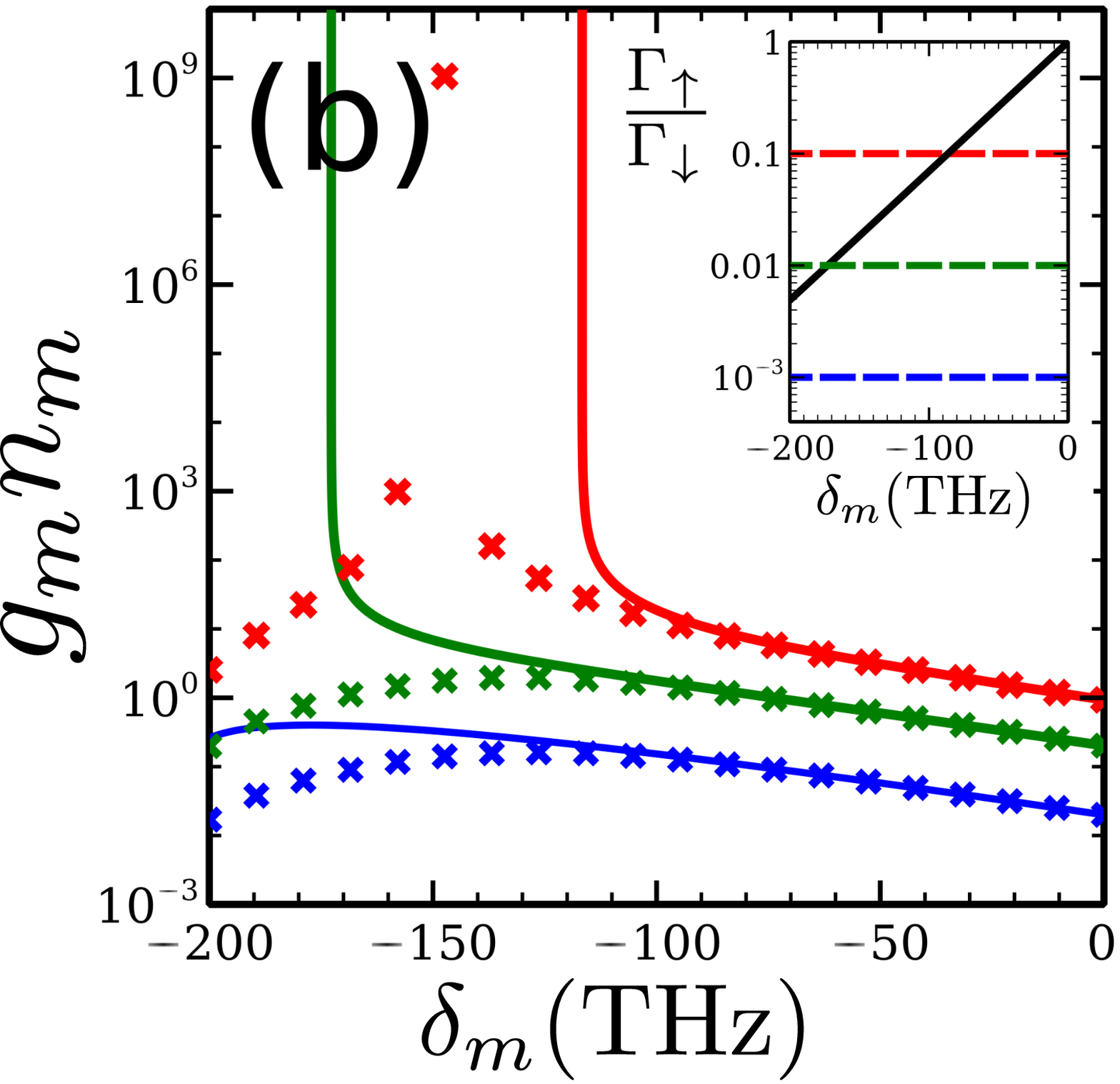}
\caption{(Color online) Mode populations $g_\modelabel n_\modelabel$ vs
  detuning $\delta_\modelabel$ for various pump strengths.  Crosses are
  results of the nonequilibrium model, and lines show Bose-Einstein
  distributions fitted to the tail of the numerical results.  Insets
  show the pump powers, $\Gu/\Gd = \text{e}^{\beta \mu_{\text{eff}}}$,
  (dashed lines) compared to the ratio
  $\Gamma(-\delta_\modelabel)/\Gamma(\delta_\modelabel) \simeq \text{e}^{\beta
    \delta_\modelabel}$ (solid, black line) which gives a good
  approximation to the threshold. Panel (a) corresponds to
  experimental losses, $\kappa=10$MHz. Panel (b) shows $\kappa=5$GHz where losses prevent thermalization. Other parameters
  are: $\gamma=100$THz, $\Gd=1$GHz, $S=0.5$, $\Omega=1$THz,
  $N=10^{11}$, $g=0.1$GHz, $T=300$K, $\delta_0=-200$THz, and the mode
  spacing $\epsilon=10$THz.}\label{fig2}
\end{figure}

In Fig.\ \ref{fig2}(a) we present results including losses for
parameter values (given in the caption) typical of those in the
experiments of Refs.~\cite{Klaers2010b, Klaers2010c, Klaers2011a}. We
fit Bose-Einstein distributions to the data by tuning the chemical
potential so that the tail matches the numerical results. These distributions agree very closely with the numerical results, even in the presence of losses.
The profile changes dramatically when the cavity loss rate is
increased, as shown in Fig.\ \ref{fig2}(b). In this case the losses
for the lowest energy modes exceeds the gain,
controlled by the rate $\Gamma(-\delta_0)$, thus preventing these modes from reaching
thermal equilibrium. As can be clearly seen in the figure, the modes
with higher energy (i.e.\  $\delta_\modelabel\simeq 0$) still match the
Bose-Einstein distributions well. At sufficiently strong pumping there
is a threshold (at much higher power than would be required if in
thermal equilibrium) above which we find a macroscopic peak appears in
an excited mode of the cavity. This mode is determined
by the lowest mode such that $\Gamma(-\delta_\modelabel)$ is large enough
to overcome the losses. For the parameters of Fig.\ \ref{fig2}(b) the thermalization still plays a role in the rates of
emission and absorption, so that it is not the mode with peak emission
rate (near $\delta_\modelabel=0$) which becomes macroscopically occupied.  However, at yet
higher decay rates, the behavior crosses over toward such 
``standard'' laser behavior, and all thermal properties are lost.
Similarly, as $T\to0$ the emission and absorption spectra
become a narrow Lorentzian peak centered at $\delta_\modelabel=0$, and
the lasing mode moves to the center of the gain peak.

Since the origin of the destruction of thermalization is the
competition between loss $\kappa$ and emission rate
$\Gamma(-\delta_0)$, it is also clear that lowering the cavity cutoff
(making $\delta_0$ more negative) has a similar effect to increasing the losses. Equilibrium behavior can only be seen when the cavity cutoff is
sufficiently close to the molecular frequency, ensuring that
$\Gamma(-\delta_0)$ is sufficiently large.

\begin{figure}
 \centering
\includegraphics[width=2.8in]{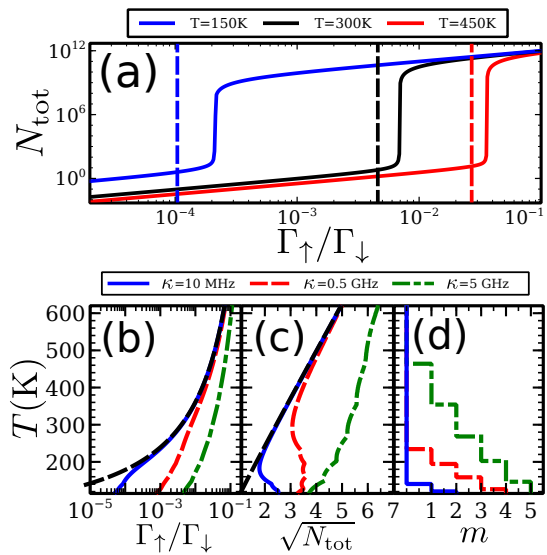}
\caption{(Color online). (a) Total number of photons in the
  cavity $N_{\rm tot}$ vs pump power. The dashed vertical lines show
  the threshold [defined by $\max(n_{\modelabel})=T/\epsilon$]. (b)
  Threshold pump power vs temperature, for various cavity loss rates
  as indicated.  The dashed (black) lines show the same threshold
  calculated for the equilibrium theory.  (c) Population
  $N_{\rm tot}$ at threshold vs temperature for the same loss rates
  shown in (b). The dashed (black) line is the equilibrium prediction
  of critical density. (d) Index of the mode which gains a macroscopic occupation for the same loss rates as in (b).  All other parameters are the same as in Fig.\
  \ref{fig2}.}\label{fig3}
\end{figure}

In order to explore the degree of thermalization as a function of
temperature and loss rates, we next consider the behavior at
threshold.  We examine two different aspects: the threshold pump power
(a measure typical when considering lasing), and the total photon
density at threshold (a measure typical for an equilibrium
condensation transition).  Figure \ref{fig3}(a) shows the total number
of photons in the cavity, $N_{\rm tot}=\sum_\modelabel g_\modelabel
n_\modelabel$, as a function of the pump rate, $\Gu$, at various
temperatures.  As expected, increasing the temperature of the phonons
reduces the asymmetry between $\Gamma(\delta_\modelabel)$ and
$\Gamma(-\delta_\modelabel)$, and thus both the pump power and total
density at threshold increase.  To explore temperature dependence we
identify the threshold as the lowest pump power where
$\max(n_\modelabel) > T/\epsilon$, with $\epsilon$ the mode spacing---note that this maximally occupied mode is not necessarily the
lowest energy mode  \footnote{This definition corresponds to
expanding around the equilibrium thermodynamic
limit~\cite{Bagnato1991,Hadzibabic2008} to next-to-leading-order in
the small parameter $\epsilon/T$.}. 

Figure \ref{fig3}(b) shows the threshold pump power vs temperature at
various cavity loss rates.  As $T$ decreases the absorption and
emission rates at a given detuning $\delta_\modelabel$ decrease. At high temperatures and small losses $\kappa$ we see very good agreement with the equilibrium prediction $\Gu/\Gd={\rm e}^{\beta \delta_0}/(1+\beta\epsilon)$.  At
low temperatures the loss rates exceed the gain for the
lowest frequencies, and thermalization breaks down.  The temperature
at which this happens increases with increasing $\kappa$.  For the
strongly lossy case, shown by the dot-dashed (green) curve
[corresponding to the $\kappa$ used in in Fig.\ \ref{fig2}(b)] the threshold remains significantly higher than the equilibrium limit  across the whole temperature range shown.  In
the high temperature limit the emission and absorption rates are
symmetric, so the threshold pump strength eventually becomes that
required to reach inversion.

A more common description of the ``threshold'' for equilibrium
condensation is the temperature dependent critical number of
particles, i.e.\  the total number of photons in the cavity at
threshold.  For particles confined in 2D one expects $N_{\rm
  tot}  = {\pi^2T^2}/{6\epsilon^2}$~\cite{pitaevskii03}.  In Fig.\ \ref{fig3}(c) we
plot this critical number for the same values of $\kappa$ as in
Fig.\ \ref{fig3}(b). Alongside these results we also plot, as the
dashed (black) curve, the equilibrium result as above. 

For small cavity losses we see that, for temperatures above $\sim200$K,
the agreement between the equilibrium results and the numerics
is very close. Below this temperature, the mode which gains a macroscopic occupation is no
longer the lowest energy mode, and the system requires stronger
pumping to go past the threshold. This leads to an increase in the critical number of photons
at low temperatures.  As $\kappa$ is increased we again see that the
temperature above which the results match a Bose distribution
increases.  There are also notable kinks in the critical number
at low temperatures.  These occur as the mode which gains a macroscopic
population jumps to higher and higher energy. To illustrate this we show in Fig.\ \ref{fig3}(d) how the index of the mode with the largest occupation at threshold varies with temperature. We see clearly that the kinks in the photon number correspond to jumps in the maximally occupied mode and that the regions which agree with the equilibrium theory only occur when the ground state has the largest occupation.

In conclusion, we have presented a simple nonequilibrium model which
accurately describes the steady state properties of the dye filled
cavity systems used to observe condensation of photons. We found that, for
relevant parameters, our model accurately predicts the transition to a condensed phase
and the equilibrium dependence of pump power and critical photon
number on temperature.  If the losses from the cavity are increased,
the temperature reduced, or the detuning increased compared to those used in
the experiment, then a crossover occurs toward behavior more typical of a
laser, and thermalization is suppressed.  These results show that, as
for polariton condensation~\cite{Szymanska2006,*Keeling2013}, a smooth
crossover between typical laser behavior and equilibrium
condensation can arise.    Future studies of the time
dynamics of how coherence arises, and the thermal distribution emerges
following the switch-on of the pump can help to clarify this
behavior, and can be predicted using the model presented here.
  
\begin{acknowledgments}
  The authors would like to thank M.\ Weitz and J.\ Klaers for useful
  discussions and acknowledge financial support from EPSRC program
  ``TOPNES'' (EP/I031014/1) and EPSRC (EP/G004714/2)
\end{acknowledgments}
 
\bibliographystyle{apsrev4-1}
%\bibliography{PhotonCondensates}


\end{document}